\documentclass[10pt,preprint2]{aastex}
\usepackage{graphics}

\def\cmm2{\,{\rm cm}^{-2}}

\newcommand{\kms}{\, {\rm km \, s}^{-1} }
\newcommand{\Mpc}{\, {\rm Mpc} }
\newcommand{\ergs}{\, {\rm ergs \, s}^{-1} }

\begin{document}

%\slugcomment{To be submitted to ApJ}
\shorttitle{XMM-Newton observations of the Perseus Cluster}
\shortauthors{Churazov et al.}

\title{XMM-Newton observations of the Perseus Cluster I: The
temperature and surface brightness structure}

\author{E. Churazov\altaffilmark{1,2}, W. Forman\altaffilmark{3},
C. Jones\altaffilmark{3}, H. B\"{o}hringer\altaffilmark{4}} 

\altaffiltext{1}{MPI f\"{u}r Astrophysik, Karl-Schwarzschild-Strasse 1, 85740
Garching, Germany}
\altaffiltext{2}{Space Research Institute (IKI), Profsoyuznaya 84/32,
Moscow 117810, Russia} 
\altaffiltext{3}{Harvard-Smithsonian Center for Astrophysics, 60 Garden St.,
Cambridge, MA 02138}
\altaffiltext{4}{MPI f\"{u}r Extraterrestrische Physik, P.O. Box 1603, 85740
Garching, Germany} 

%\begin{document}  originally here
%\maketitle

\begin{abstract}
We present preliminary results of the XMM-Newton 50~ksec observation
of the Perseus cluster which provides an unprecedented view of the
central 0.5 Mpc region. The projected gas temperature declines
smoothly by a factor of 2 from a maximum value of $\sim$ 7~keV in the
outer regions to just above 3~keV at the cluster center. Over this
same range, the heavy element abundance rises slowly from 0.4 solar to
0.5 solar as the radius decreases from $14'$ to $5'$, and then rises
to a peak of almost 0.7 solar at $1.25'$ before declining to 0.4 at
the center.  The global east/west asymmetry of the gas temperature and
surface brightness distributions, approximately aligned with the chain
of bright galaxies, suggests an ongoing merger, although the modest
degree of the observed asymmetry certainly excludes a major merger
interpretation. The chain of galaxies probably traces the filament
along which accretion has started some time ago and is continuing at
the present time. A cold and dense (low entropy) cluster core like
Perseus is probably well ``protected'' against the penetration of the
gas of infalling groups and poor clusters whereas in non-cooling core
clusters like Coma and A1367, infalling subclusters can penetrate
deeply into the core region. In Perseus, gas associated with infalling
groups may be stripped completely at the outskirts of the main cluster
and only compression waves (shocks) may reach the central regions. We
argue, and show supporting simulations, that the passage of such a
wave(s) can qualitatively explain the overall horseshoe shaped
appearance of the gas temperature map (the hot horseshoe surrounds the
colder, low entropy core) as well as other features of the Perseus
cluster core. These simulations also show that as compression waves
traverse the cluster core, they can induce oscillatory motion of the
cluster gas which can generate multiple sharp ``edges'', on opposite
sides or the central galaxy. Gas motions induced by mergers may be a
natural way to explain the high frequency of ``edges'' seen in
clusters with cooling cores.

\end{abstract}


%\label{firstpage}

%\sloppypar

\section{Introduction}
The Perseus cluster of galaxies (Abell 426) is the X-ray brightest
nearby cluster (with $z=0.01756$, $1'$ corresponds to $\sim$30 kpc
(29.7 kpc) for $H_0=50~\kms~\Mpc^{-1}$ which we use throughout). It
was first reported as an extended X-ray source from {\em UHURU}
observations (Forman et al. 1972) and first images (Gorenstein et
al. 1978; see also Fabian et al. 1974; Wolff et al. 1974; Malina et
al. 1978) provided a glimpse of the X-ray structure of Perseus.
Detailed X--ray images were obtained with the Einstein IPC
(Branduardi--Raymont et al. 1981) and HRI (Fabian et al. 1981), the
ROSAT PSPC (Schwarz et al. 1992, Ettori, Fabian, White 1998), HRI
(B\"{o}hringer et al. 1993) and recently with Chandra (Fabian et
al. 2000, Schmidt, Fabian \& Sanders, 2002). The cluster has a
prominent X--ray surface brightness peak centered on the active galaxy
NGC~1275, containing a strong core-dominated radio source (Per~A, 3C84)
surrounded by a lower surface brightness halo (e.g. Pedlar et
al. 1990, Sijbring 1993). Many studies have explored the correlations
between X-ray, radio, optical, and ultraviolet emission in the core
(see e.g. McNamara, O'Connell \& Sarazin 1996 and references
therein). The proximity and brightness of the cluster make it a
natural target for studies of a relatively relaxed cluster with a cold
core. In this contribution, we report preliminary results of the
XMM-Newton 50~ksec observations of the Perseus cluster.

The structure of the paper is as follows: in Section 2 we discuss XMM-Newton
observations of the cluster, in Section 3 we describe the analysis of the
data and present the results, in Section 4 we discuss possible implications
of the results, and in Section 5 we summarize our findings. 
The analysis of the substructure in the very core (less than $2'$) of
the Perseus cluster and details of the heavy element abundance and
distribution are beyond the scope of this paper and will be reported
in subsequent publications.

\begin{figure}[ht]
\plotone{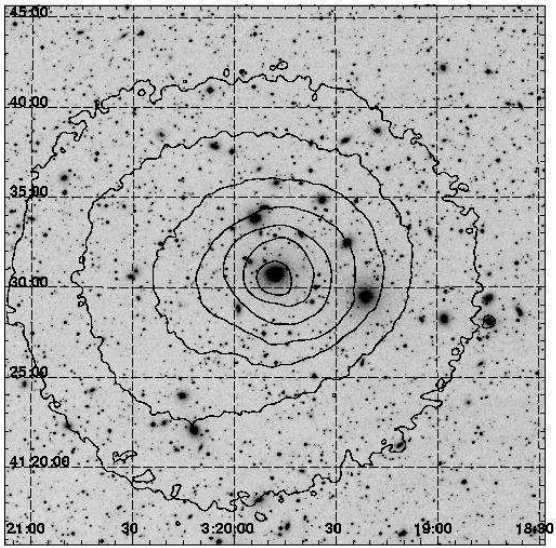}
\caption{Surface brightness contours in the 0.3--5 keV
energy range (corrected for the background and telescope vignetting) superposed
onto a DSS optical image ($30' \times 30'$). The X--ray image has been
smoothed with an $8''$ Gaussian. The central galaxy is NGC~1275. A
chain of bright galaxies is visible to the West. }
\label{fig:image}
\end{figure}

\section{Observations and background subtraction}
\begin{figure}[ht]
\plotone{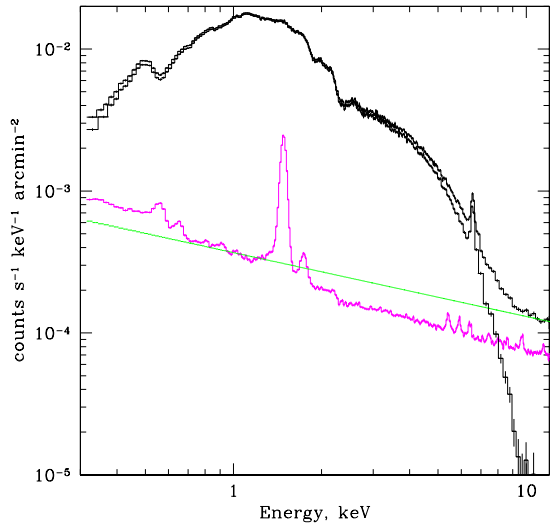}
\caption{The spectrum of the $8'-13.75'$ annulus around NGC~1275 and the
background spectrum. Two lower curves show the blank field
``quiescent'' background and the additional ``flare'' background component
(power law) assumed in the analysis. Two upper curves show the
source spectrum with the quiescent background subtracted (upper curve)
and the source spectrum with both the quiescent and flare background
components subtracted (lower thick curve). The normalization of the
flare background has been estimated using the data above $\sim$11 keV.
\label{fig:bckg}}
\end{figure}

The core of the Perseus cluster was observed with XMM-Newton on 2001,
Jan. 30 with a total integration time of $\sim$53 ksec (observation ID
\#00851101). The pointing was centered on NGC~1275
(Fig.~\ref{fig:image}). We report below on the results obtained with
the MOS instrument only. Results from the RGS and PN observations will
be reported elsewhere.  Calibrated event lists were generated using
SAS v5.3.  As is known the MOS background has a steady (quiescent)
component partly associated with cosmic X--ray background and variable
component due to charged particles (e.g. Lumb et al. 2002).  To check
the level of the variable background component, we used the same
tracer as implemented in the SAS task {\em emchain} -- i.e. the count
rate of events with energy deposited in a single pixel higher than a
threshold value ({\em ``REJECTED\_BY\_GATTI''} -- in SAS notation).
Although there are no obvious spikes in the light curve of such
events, the count rate is several times higher than the recommended
threshold value for the entire observation, indicating a significantly
enhanced background level. Thus a straightforward application of the
recommended cleaning procedure (using {\em emchain} default settings)
would remove essentially all the data.  To use the dataset, we had to
adopt special procedures to correct for the variable background
component. As a first step, the steady component of the background was
subtracted using data from blank field observations, provided by the
XMM SOC site (background data files are from 2002, Jan., see Lumb et
al. 2002). The spectra of all but the central chip were accumulated
for MOS1 and MOS2. Assuming that the variable component of the
background has a power law shape (in counts/channel space) with the
slope of $B(E)=A\times E^\alpha$ and using the data in the 11-12 keV
range, we calculated the normalization of the variable component. The
slope $\alpha$ of the variable background spectrum is known to vary
from observation to observation and (rather arbitrarily) has been
fixed at $\alpha=-0.45$. We then assumed that the intensity of the
variable component per unit area is constant across the detector and
corrected all observed spectra for this variable component. The sample
source and background spectra for the $8'-13.75'$ annulus is shown in
Fig.~\ref{fig:bckg}. The same procedure applied to the publicly
available datasets with the ``quiescent'' background typically yields
a normalization for the additional background component that is at
least an order of magnitude lower. 

\begin{figure}[ht]
\plotone{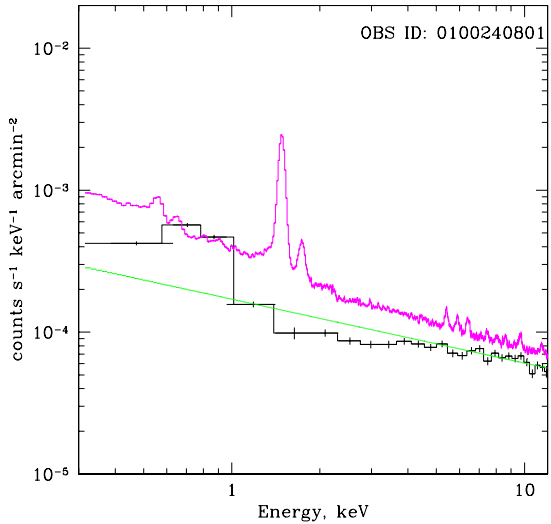}
\caption{The same as in the previous figure but for the observation performed
$\sim$14 hours after the end of the Perseus observation. The
background level was still enchanced compared to the quiescent state,
although a factor of $\sim$2 lower than in the Perseus observation. The
upper curve shows the quiescent background spectrum. The histogram is
the spectrum accumulated over the $8'-13.75'$ annulus and corrected
for the quiescent background. The power law shows the predicted
``flare'' background component, obtained with the same procedure as
was used for the Perseus observation.
}
\label{fig:buz}
\end{figure}

To verify the quality of the background subtraction procedure we used
a publicly available 30 ksec observation (OBSID 0100240801) performed
$\sim$14 hours after the end of the Perseus observation. In this
observation the background level was still enchanced compared to the
quiescent level, although a factor of $\sim$2 lower than in the
Perseus observation. There are no strong sources in this field (apart
from the target - a compact source) and it is easier to see how well
the background is corrected in the regions far from the center of the
field of view. For this data set we repeated exactly the same
background correction procedure as for the Perseus observation. The
results are shown in Fig.~\ref{fig:buz}. One can see that above
$\sim$4 keV the predicted extra background component matches almost
perfectly the accumulated spectrum (corrected for quiescent
background). Below ~2 keV some residuals are present, but given the
brightness of Perseus at these energies (factor 10 to 100 brighter
than the background) such residuals have no significant impact on the
results of the spectral fits.

The procedure and functions described above are the result of
extensive tests with various shapes of the additional background and
different methods to determine its normalization.  The results
reported below were found to be robust to the remaining uncertainties
in the background subtraction.  Further refinement of the background
correction procedure may change slightly some of the numerical values,
in particular the maximum gas temperature close to the edge of the
field of view by $\sim$5-10\%, but cannot affect any of the principle
results in the paper. More subtle features, like the limits on the
possible nonthermal emission components, which are in principle
attainable with the required statistics are deferred to future
work. For the same reason we report only on the MOS observation, for
which we achieved the most robust results.

For the subsequent analysis we used MOS data with patterns in the
range 0-12 and the recommended value of the flag (XMM\_EA). For the
spectral analysis we use one of the MOS response matrixes provided by
the XMM SOC (namely ``m1\_thin1v9q19t5r5\_all\_15.rsp'') and assume
that the same response (corrected for energy dependent vignetting) is
applicable for all regions.

\section{Analysis}
\begin{figure}[ht]
\plotone{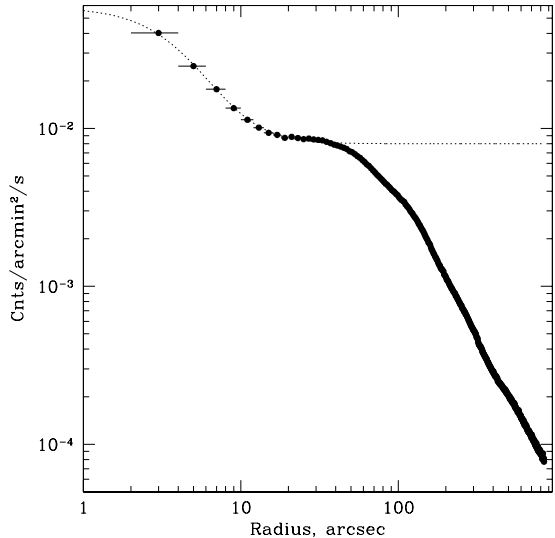}
\caption{Radial dependence of the surface brightness in the 0.3--5
keV band. The dotted line shows the sum of the telescope Point Spread
Function (PSF) plus a constant, illustrating that the central excess is
consistent with being due to the NGC~1275 nucleus contribution.
\label{fig:sb}}
\end{figure}

Although the cluster is not perfectly symmetric (see Fig.~\ref{fig:image})
it is illustrative to derive the radial dependence of major
parameters, in particular the density, temperature and heavy element abundance, assuming
spherical symmetry with the center at NGC~1275. 

The radial dependence of the surface brightness is shown in
Fig.~\ref{fig:sb}. The dotted line shows the
expected profile (taking into account the PSF of the telescope)
for a point source plus constant surface brightness. The
central excess (within $10''$) is therefore consistent with being due to
the contribution of the unresolved,  compact source -- the nucleus of NGC~1275.   

\subsection{Spectrum of NGC~1275}
\begin{figure}[ht]
\plotone{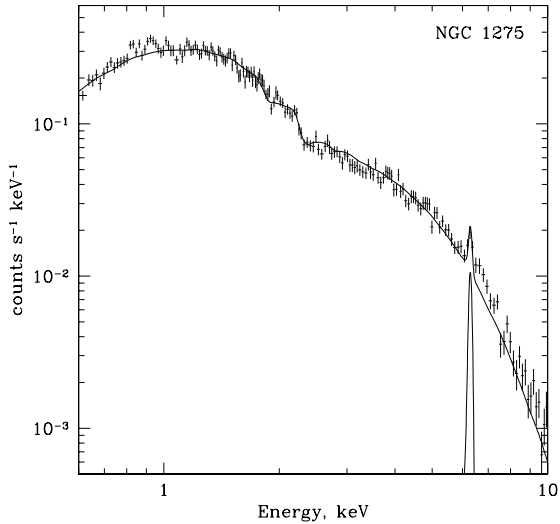}
\caption{NGC~1275 spectrum determined as the difference between
the spectra extracted from the $14''$ (radius) circle and the annulus
from $14''$ to $28''$. The position of the line is consistent with the
neutral iron fluorescent line at 6.4 keV with the redshift of 0.0176.
\label{fig:specn}}
\end{figure}

Given the XMM-Newton spatial resolution and complicated
surface brightness distribution near the cluster core, the precise
determination of the NGC~1275 nuclear spectrum is difficult. We used the
central $14''$ (radius) circle and the $14''$--$28''$ annulus as the
source and background  regions respectively. The resulting spectrum is shown in
Fig.~\ref{fig:specn}. It is fairly well fit with an absorbed 
($N_H=1.2\times10^{21} \cmm2$) power law with a photon index of 1.65 and
a narrow line at an energy of 6.289 keV (with  $1\sigma$
confidence interval from 6.276 to  6.313 keV) and an equivalent width of
$\sim$165 eV. The assumption that the line is due to fluorescent emission
of neutral iron at 6.4 keV implies a redshift of $z=0.0176\pm0.003$,
fully consistent with the optically determined redshift of NGC~1275
$z=0.01756$. A significant amount of cold gas and dust is known to be
associated with the high velocity (HV) system (Minkowski 1957, see
also Conselice, Gallagher, \& Wyse 2001), but the excellent agreement
between the 
redshifts of the line and the galaxy indicates that reprocessing by
the HV gas does not contribute significantly to the observed spectrum. The 6.7
keV line which is dominant in the cluster gas emission is very weak in
the nuclear spectrum, suggesting that background 
has been subtracted with sufficient accuracy. We conclude that the observed
spectrum of the nucleus is consistent with typical AGN
spectra. The luminosity of the nucleus derived from the best fit
parameters is of the  order of $10^{43}\ergs$ in the 0.5--8 keV range
and is subject to systematic uncertainties (at the level of 20\%) due
to background subtraction and PSF correction. 
 
\subsection{Projected parameters}

\begin{figure}[ht]
\plotone{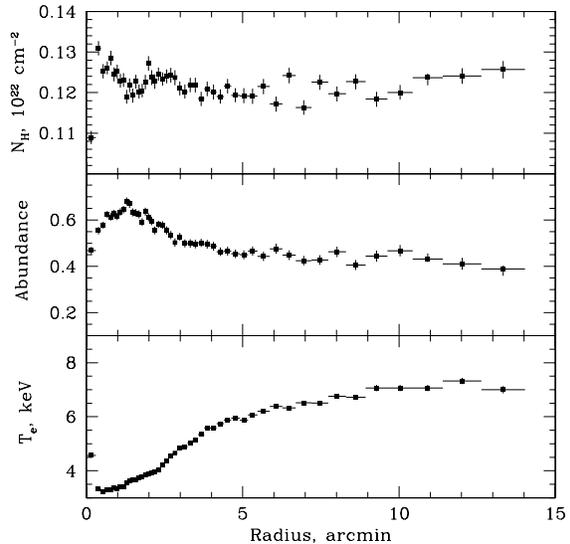}
\caption{Radial profiles of $N_H, T_e$ and heavy element abundance
calculated for a set of annuli centered at NGC~1275. The parameters were
obtained using a single temperature APEC model. The first two data
points in the innermost region are affected by the presence of the
bright compact source -- the nucleus of NGC~1275.
\label{fig:proj}}
\end{figure}
Spectra were accumulated in a set of annuli centered at NGC~1275. The
resulting spectra were fit (in the 0.5--9 keV range) with a
single temperature APEC (Smith et al. 2001) model in XSPEC
(version 11.2.0; Arnaud 1996) with the gas temperature 
$T_e$, abundance of heavy elements $Z$ (relative to the solar values of
Anders and Grevesse 1989) and low energy photoelectric absorption column
density $N_H$ as free parameters. The redshift 
has been fixed at the optically determined value for NGC~1275:
$z=0.01756$ (Strauss et al. 1992). The projected radial dependence of all parameters is
shown in Fig.~\ref{fig:proj}. The derived value of $N_H\sim
1.2-1.3\times10^{21}$ cm$^{-2}$ is comparable with (and even somewhat
lower than) 
the galactic hydrogen column density in this direction $N_H\sim
1.5\times10^{21}$~$cm^{-2}$ (Dickey \& Lockman 1990). No intrinsic
absorption is required by the data and no large variations are
observed with radius. However we note that, given the uncertainty in the
background subtraction procedure described above, it is possible that
we systematically underestimate the absorbing column density. 
The temperature shows a clear decrease from 7 keV at large radii to
3.7 keV near the 
center of the cluster, except for the very central region which is
contaminated by 
hard emission from the NGC~1275 nucleus. The abundance of heavy
elements shows an opposite trend, decreasing (outside the central $1'-2'$
region) with distance from the center. 

\subsection{Asymmetric surface brightness substructure}
\begin{figure*}[ht]
\plotone{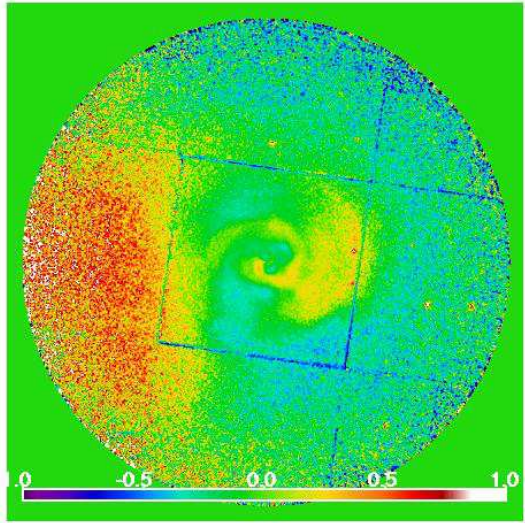}
\caption{Relative deviation of the surface brightness from the
azimuthally averaged value. The image shown is $30' \times 30'$.
Clearly seen are: the region of excess emission
to the East from the center, an edge-like structure $5'-7'$ to the West from the
center and complicated structure in the inner $1'-2'$ region. 
\label{fig:divs}}
\end{figure*}
As is clear from Fig.~\ref{fig:image} and known from earlier Einstein
and ROSAT observations (Branduardi-Raymont et al. 1981),  the surface brightness is not
symmetric around NGC~1275. In order to further enhance the asymmetric
structure, we plot in Fig.~\ref{fig:divs} the relative deviations of the surface
brightness from the azimuthally averaged surface brightness value
(i.e. $[Data-Model]/Model$). The resulting image has been smoothed
with a $2''$ Gaussian. Apart from the complicated structure in
the innermost ($\sim$2$'$) region probably dominated by the interaction
of the thermal gas with the central AGN radio lobes (e.g. B\"ohringer et
al., 1993, Churazov et al. 2000, Fabian et al. 2000, 2002) there are two
regions of prominent excess emission on large scales -- one
$\sim$4$'$-7$'$ west of
the center and one $\ge5'$ east of the center. The same
structures were also clearly seen in earlier ROSAT PSPC and HRI
images (e.g. Schwarz et al. 1992). The possible nature of these structures is
further discussed in Section 4. 

\subsection{Deprojection analysis}
The above radial dependencies of the gas parameters are affected by
projection effects and the azimuthal asymmetry of the cluster. The
first problem can be solved using deprojection analysis. Various
flavors of deprojection techniques are frequently used (e.g. Fabian et
al. 1981, Kriss, Cioffi \& Canizares 1983, McLaughlin 1999). In our
analysis we assume spherical symmetry but make no specific assumption
about the form of the underlying gravitational potential. We first
calculate the surface brightness (in a given energy band) in a set of
$n_a$ annuli and choose a set of $n_s$ spherical shells. The radii of
the annuli and shells need not be the same. The gas parameters are
assumed to be uniform inside each shell. The outermost radii of the
annuli and shells are chosen so that $r(n_s)<r(n_a)$. The emissivity
of the gas for radii larger than $r(n_s)$ is assumed to decline with
the radius as a power law with a given slope, namely
$\mathcal{E}=\mathcal{E}_0 r^{-4}$. The normalization of
this extra component $\mathcal{E}_0$ is an additional free parameter
of the model.  One can then write a simple expression which describes
the contribution of all shells and outer regions from $r(n_s)$ to
infinity to the surface brightness $S(j)$ in a given annulus $j$.
\begin{eqnarray}
S(j)=\sum_{i=1,n_s} A(i,j)\mathcal{E}(i)+B(j)\mathcal{E}_0,
\end{eqnarray}
where $\mathcal{E}(i)$ is the emissivity of a given shell. 
A simple analytical expression for $A(i,j)$ and $B(j)$ which are
functions of the radii of corresponding shell and annulus can be easily written
(e.g. Kriss, Cioffi and Canizares 1983, McLaughlin 1999). 

We then formulate a least squares problem -- what set of
emissivities in our set of shells will provide the best description of the
surface brightness recorded in our set of annuli:
\begin{eqnarray}
\sum_{j}\left [\sum_{i} A(i,j)\mathcal{E}(i)+B(j)\mathcal{E}_0 - S(j)
\right ]^2 /\sigma_j^2 \nonumber \\ = min,
\end{eqnarray}
where $\sigma_j$ is the statistical error associated with the surface
brightness in a given annulus. As usual the differentiation of this
relation with respect to $S(j)$ and $\mathcal{E}_0$
yields a system of linear equations, which can easily
be inverted. The properties of the inverse matrix $C(j,i)$ (in particular
an error enhancement due to ill-conditioned problem) can be easily
controlled by making the spherical shells broader. The emissivities of
the shells are then expressed through the observed surface brightness
distribution as an explicit linear combination:
\begin{eqnarray}
\mathcal{E}(i)=\sum_{j=1,n_a} C(j,i) S(j)
\end{eqnarray}
\begin{figure}[ht]
\plotone{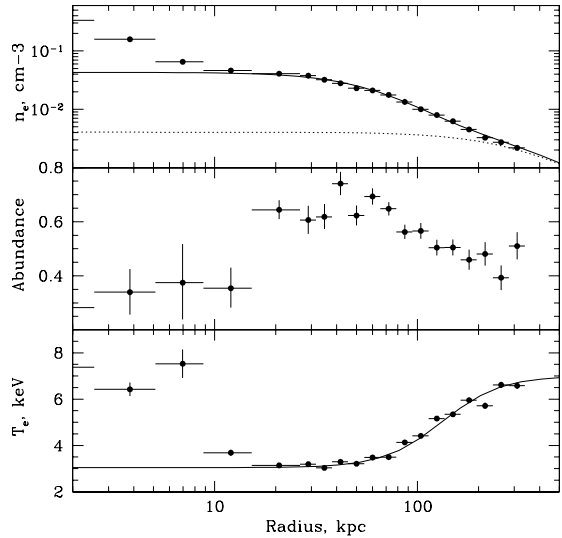}
\caption{Radial profiles of deprojected parameters. The parameters were
obtained using a single temperature APEC model with fixed low energy
absorption fitted to deprojected spectra for a set of spherical shells
centered at NGC~1275. The data points in the innermost region ($\sim$10
kpc) are affected by the presence of the bright compact source -- the
nucleus of NGC~1275. Analytical approximations for the density and 
temperature are shown as solid lines (see text).   
\label{fig:deproj}}
\end{figure}

We then accumulate a set of spectra (corrected for the background and
vignetting) for each annulus. The emission spectrum associated with each
shell can then be calculated as the same linear combination of the
observed spectra. Since the whole procedure is linear the (statistical)
errors can be propagated straightforwardly. The resulting spectra for
every shell are fit with standard models in
XSPEC. The radial dependence of the electron density, temperature and
heavy element abundance is shown in Fig.~\ref{fig:deproj}. As usual
for any kind of deprojection analysis the errors
on the deprojected values of temperature and abundance are
significantly larger than for the projected spectra. However the
biggest uncertainty in the deprojected parameters probably comes from the
assumption of spherical symmetry.  As is clear from Section 3.3 the
cluster is significantly asymmetric. Furthermore when deriving the density
in a given shell we assume that X--ray emitting gas uniformly fills the
volume of the shell. Nonetheless we present below two
simple analytical approximations to the density and
temperature distributions which crudely characterize the radial behavior
of these two parameters (shown in Fig.~\ref{fig:deproj} with the solid lines):  
\begin{eqnarray}
n_e=\frac{3.9\times10^{-2}}{[1+(\frac{r}{80})^2]^{1.8}}+
\frac{4.05\times10^{-3}}{[1+(\frac{r}{280})^2]^{0.87}}~~~{\rm cm}^{-3}
\label{ne}
\end{eqnarray}
where the second term is the density distribution on larger scales
taken from Jones \& Forman (1999) and shown in Fig.~\ref{fig:deproj}
with the dotted line.
\begin{eqnarray}
T_e=7\frac{[1+(\frac{r}{100})^3]}{[2.3+(\frac{r}{100})^3]}~~~{\rm keV}
\label{te}
\end{eqnarray}

The analytical approximations shown in Fig.~\ref{fig:deproj} clearly
fail to follow the data in the innermost region (within the central
$\sim$10 kpc). Given the XMM-Newton angular resolution, the major
problem here is the contribution from the bright compact source -- the
nucleus of NGC~1275.  This makes Chandra much more suitable for
studies of the innermost region.

Schmidt et al. (2002), using Chandra data and excluding the
contribution from the nucleus, showed that within about 50 kpc, the
deprojected temperature profile is essentially flat (see their Fig.~7)
with a mean of approximately 3.1~keV in excellent agreement with the
XMM-Newton temperature values from 10-50~kpc (away from the region in
which the nuclear source contributes to the XMM-Newton spectral
data). The temperature profile derived from the projected Chandra
spectra in finer radial bins (Fig.~2 from Schmidt et al. 2002) does
however show a slight increase in temperature in the innermost ($<$10
kpc) bin. It is possible that projection effects (due to the presence of X--ray
``holes'' in the inner region) increase somewhat the apparent gas
temperature there.  Therefore, equation~\ref{te}, with its
essentially constant temperature at $r<50$ kpc is a reasonable
approximation. A comparison of the azimuthly averaged Chandra surface
brightness profile in the inner region with the XMM-Newton data also
shows that a gas density profile with a flat core (eq.~\ref{ne}) is a
reasonable approximation to the data.

\subsection{Projected Temperature Structure}

\begin{figure*}[ht]
\plotone{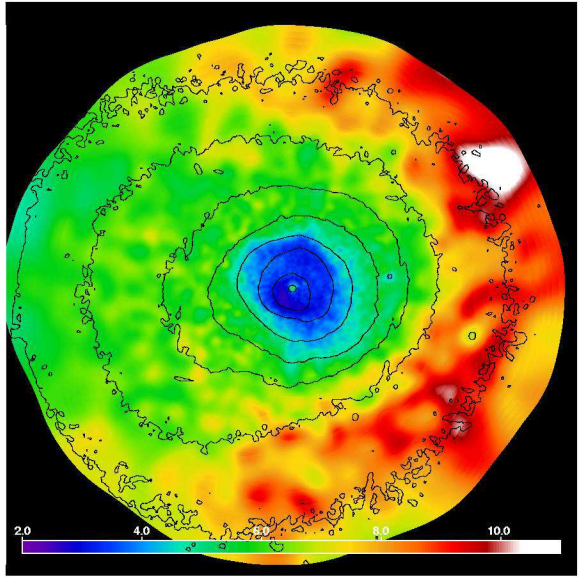}
\caption{Gas temperature map ($30'\times 30'$). The temperature is
derived from adaptively smoothed data with an effective number of
counts within each smoothing window of $\sim 10^4$. Contours show the
surface brightness distribution. 
\label{fig:tmap}}
\end{figure*}

To calculate the projected gas temperature distribution, we employed the
method described in Churazov et al. (1996, 1999) and applied previously to the
ASCA data (e.g. Donnelly et al. 1998, 2001). Briefly we use template spectra
corresponding to emission from an optically thin plasma (convolved
with the MOS
energy response) with temperatures 3 and
9 keV with a given metallicity and low energy photoelectric absorption and
determine the best fit weights of these template spectra needed to describe
the spectrum $S(E)$ observed in a given pixel of the image, i.e.
\begin{eqnarray}
S(E)=A\cdot M(T_1,E)+B\cdot M(T_2,E),
\end{eqnarray}
where $M(T,E)$ is the template spectrum for a given value of
temperature $T$. The temperature is then
calculated as a function of the relative weights of the template spectra.
As is shown in Churazov et al. (1996) an 
expectation value of the temperature calculated this way is usually very close
(within a few percent) to the temperature obtained by conventional
spectral fitting under the assumption of a single temperature plasma with fixed
metallicity and absorption. Compared to hardness ratios, typically used for
the same purpose, this method is significantly more sensitive. The
determination of weights is a linear procedure (i.e. fast) and further
smoothing can be applied after the images containing weights are calculated.
The resulting temperature map for Perseus is shown in Fig.~\ref{fig:tmap}. An adaptive
smoothing has been applied to the map in such a way that each value of
the temperature has been calculated using regions containing 
$\sim 10^4$ counts. Comparison with the results of direct
spectral fitting of individual regions shows good agreement with the overall
temperature structure.

Compared with the previous ASCA temperature maps (Arnaud et al. 1994,
Furusho et al. 2001) XMM-Newton data show the innermost region in much
greater detail, while the global asymmetry is broadly consistent with
the lower resolution ASCA maps, which cover a much larger area. The temperature
structure of the very central region (within $1'-2'$) is consistent with the
recent Chandra results (Fabian et al. 2000, Schmidt et al. 2002).

On the spatial scales of 1--15 arcminutes, which are particularly well
mapped by the XMM-Newton observations, the most remarkable feature of the
temperature map is a horseshoe-shaped high temperature region to the
West, enveloping the central cooler region. Thus the global east-west
asymmetry of the cluster seen in the surface brightness distribution
(Fig.~\ref{fig:divs}) is also clearly present in the gas temperature
distribution.

\begin{figure}[ht]
\plotone{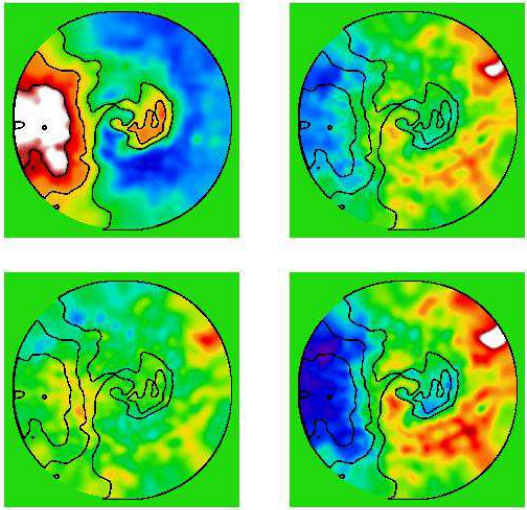}
\caption{Relative deviations from azimuthly averaged values of a)
surface brightness $I$ b) temperature $T$ c) pressure $p=T\sqrt{I}$
and d) entropy $s=T/I^{1/3}$. Contours mark the regions of excess surface
brightness (from the first image). The color scale is the same for all
three plots (from -0.5 to +0.5, see color bar in Fig.7).
Clearly the azimuthal variations in
pressure are smaller than in any other quantity.}
\label{fig:press}
\end{figure}

If the cluster gas is in a state of a hydrostatic equilibrium then the
gas density and temperature should remain constant along the
equipotential surfaces set by the gravitating mass. In the example of
a spherical or elliptical potential, all features in the observed
surface brightness or temperature distribution should reflect the
symmetry of the underlying potential. Shocks or strong sound waves
(e.g. produced by a merger) may reveal themselves as distinct features
in the observed images or temperature maps. The variations of density
and temperature in shocks are obviously correlated leading to a
positive correlation of the variations in surface brightness images
and temperature maps. The observed variations (Fig.~\ref{fig:divs} and
Fig.~\ref{fig:tmap}), on the contrary, suggest that in Perseus the
variations of the temperature and surface brightness are
anticorrelated. One can see that all the most prominent positive
features in Fig.~\ref{fig:divs} correspond to regions of lower
(relative to the mean value at the same distance from the center)
temperature. To show this more clearly in Fig.~\ref{fig:press} we plot
side by side the deviations of the surface brightness and temperature
from the azimuthally averaged values. The surface brightness image $I$
and the temperature map $T$ can be further combined in the form of two
additional images $P=T\sqrt{I}$ and $S=T/I^{1/3}$. Since the surface
brightness scales as the square of the electron density
(i.e. $I\propto n_e^2$) then $P\propto T n_e$ and $S\propto
T/n_e^{2/3}$. Therefore these two images $P$ and $S$ can be used to
approximately characterize the azimuthal variations of pressure and
entropy\footnote{we use the term entropy for the quantity
$S=T/n_e^{2/3}$, true specific entropy is determined by the logarithm
of this quantity} of the gas. These variations are also shown in
Fig.~\ref{fig:press}.  It is clear that the relative amplitude of the
azimuthal variations in pressure is smaller than that for any of the
other quantities shown in Fig.~\ref{fig:press} and there is no clear
east-west asymmetry in the pressure distribution. This argues against
the attribution of the bulk of the observed substructure in the
surface brightness distribution and temperature to strong
shocks. Instead the observed features clearly resemble edges or ``cold
fronts'' found recently in several clusters (e.g. Markevitch et
al. 2000, 2002, Vikhlinin et al. 2001). The possible origin of this
substructure is discussed below.

\section{Discussion}
During the lifetime of any cluster, it experiences a few major mergers
with another cluster of comparable mass and much more frequent
accretion of smaller clusters or groups. A chain of bright optical
galaxies extending to the west from NGC~1275 suggests that at present
the Perseus cluster is merging with a subcluster coming from that
direction. The effect of a merger on the appearance of the cluster
obviously depends strongly on the mass ratios and gas fractions of
merging clusters (see recent reviews by Evrard \& Gioia 2002, Sarazin
2002, Schindler 2002 and references therein).  Given that the
deviations from  spherical symmetry in the surface brightness and
temperature, although clearly present, are not very large then we
probably have a case of a merger with a relatively small subcluster or
group of galaxies. Furthermore the lack of large azimuthal pressure
variations (Fig.~\ref{fig:press}) suggests that we observe the merger
in a progressed stage when strong shocks have already left the central
region of the cluster mapped with XMM-Newton.  The numerical (N-body
and fluid dynamics) simulations of the merger most relevant for the
Perseus case are probably those of Gomez et al. (2002). Their merger 7
(see Table 1 from Gomez et al., 2002) has a mass ratio 1:16 and small
gas fraction for the subcluster so that the cooling flow is not
disrupted during the merger. One can see some features in the gas
density and temperature distributions in their simulations (Fig.~5 and
6 in Gomez et al., 2002) which are qualitatively similar to the
structures found in the Perseus data.

In Perseus the most striking feature is a clear edge in the surface
brightness which separates cooler gas in the very core from the hotter
horseshoe-shaped region in the West, enveloping the core.  One
plausible explanation of these structures is a contact discontinuity
separating main cluster gas from the gas of the infalling
subcluster. The shape of the contact discontinuity depends primarily
on the density profiles of colliding clusters. For two identical
clusters involved in a head-on merger the contact discontinuity is a
plane perpendicular to the line connecting the centers of two clusters
and located half way between the centers. In the case of a merger of a
rich cluster with smaller and less dense subcluster the contact
discontinuity shifts towards the center of the smaller subcluster and
bends due to stronger impact of the ram pressure onto the outer
regions of the smaller subcluster, giving the infalling gas a conical
shape (see e.g. Fig.~5 in Gomez et al. 2002). The presence of a very dense
core in the main cluster may reverse this trend and bend the contact
discontinuity in the opposite direction so that it will envelop the very
core of the main cluster. This behavior is less obvious in the
simulations by Gomez et al. (2002) although flattening of the contact
discontinuity is definitely seen.  Assuming that merger 7 of Gomez et
al. (2002) is a close analog of the Perseus merger, one would conclude
that the present state of the Perseus cluster corresponds to a time
$\sim0.25$ Gyr after the core crossing when the contact discontinuity
is located within $\sim$ 200 kpc of the center of the main cluster
(upper right panel of their Fig.~4). After 0.5-0.75 Gyr the dense and
cold gas of the main cluster rebounds and shifts the bulk of the
subcluster gas to larger distances from the center (middle right panel
of their Fig.~4).

An alternative explanation of the edge and horseshoe structures in
Perseus assumes that the contact discontinuity is located farther away
from the cluster core and the observed structures are solely composed from
the disturbed gas of the main cluster. If the chain of bright galaxies
marks the ongoing accretion direction of  very small groups of
galaxies, it is possible that (given the richness of Perseus)
in each individual merger event the gas of the infalling
subcluster is decelerated far from the Perseus core and never comes
close to the center. Such a situation resembles the picture of a
piston which initially is moving  through the gas but
then stops. The gas in the core of the Perseus cluster would then see
a compression wave (shock) followed by a rarefaction wave. The gas
density and temperature increase during the compression phase, but
then decrease as the gas reexpands. There should be only a modest
increase in the final gas temperature associated with the increase of
the entropy in the shock. The velocity of the gas in the core of the
main cluster also first increases and then falls back to zero, with
the net effect that every volume element of gas is displaced some
distance from its initial position. Since the density, temperature and
pressure change significantly across the core (particularly in the
direction perpendicular to the direction of the compression wave
motion) the impact of the passing compression/rarefaction wave will 
differ in terms of the increase in temperature and the net
displacement. If the characteristic amplitude of
the wave propagating through the nonuniform medium scales as the square
root of the density $n_e$ then the amplitude of the initial 
displacement at a distance $r$ from the center $X(r)$ scales as
$\sqrt{\frac{1}{n_e(r)}}$. Therefore the wave would induce a smaller net
displacement in the very core region (where the density is the
highest), than in the region farther from the core. Such a process can
lead to characteristic ``horseshoe'' structures in the temperature and
density distributions that are qualitatively similar to those observed
in the Perseus cluster. However the relative amplitude of the
displacement $X(r)/r\propto \sqrt{\frac{1}{n_e(r)r^2}}$  is a
decreasing function of radius for  density profiles less steep than
$r^{-2}$. Therefore the relative amplitude of the displacement in the
very core of the cluster will still be strong. The displaced gas in
the core will then try to restore hydrostatic equilibrium and 
oscillate in the potential.

\begin{figure}[ht]
{\centering \leavevmode
\centerline{\includegraphics[width={0.4\columnwidth}]{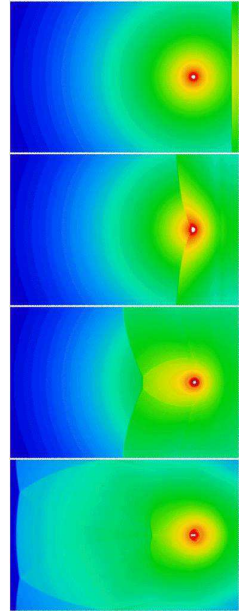}}}
\caption{Snapshots of the gas density at 0, 0.3, 0.7, 1.6 Gyr from the
beginning of the simulations. The box shown is 4$\times$6 Mpc. The
density, temperature and velocity of the gas in a narrow slab on the
right side of the box were set to mimic the impact of a piston which
is first driven into the gas at the Mach number of 2 and then suddenly
stopped. 
\label{fig:simden}}
\end{figure}

We illustrate the bending of the shock and the induced oscillatory motion of
the gas in the core with a simple numerical simulation of a plane
compression wave passing through a cluster gas. The simulations were
made using the ZEUS-2D code (Stone \& Norman 1992a, b). In our
simulations we employed an ideal gas equation of state with
$\gamma=5/3$. Equations \ref{ne} and \ref{te} were used for the
initial density and temperature distributions. The gravitational
potential was calculated from the same density and temperature
distributions and assumed to be static. The computational domain spans
a $6\times4$ Mpc region and is covered by 600 $\times$ 400 grid
points. At some distance from the cluster, we modify the gas velocity,
density and temperature in a slab 200 kpc wide, to mimic the 
situation when a piston is first driven into a cluster gas with a Mach
number of $\sim$2 and then stops.  The results
shown below were obtained from the simulations on a Cartesian grid,
while simulations on the cylindrical grid produce qualitatively very
similar results. 

\begin{figure*}[ht]
{\centering \leavevmode
\centerline{\includegraphics[width={2.0\columnwidth}]{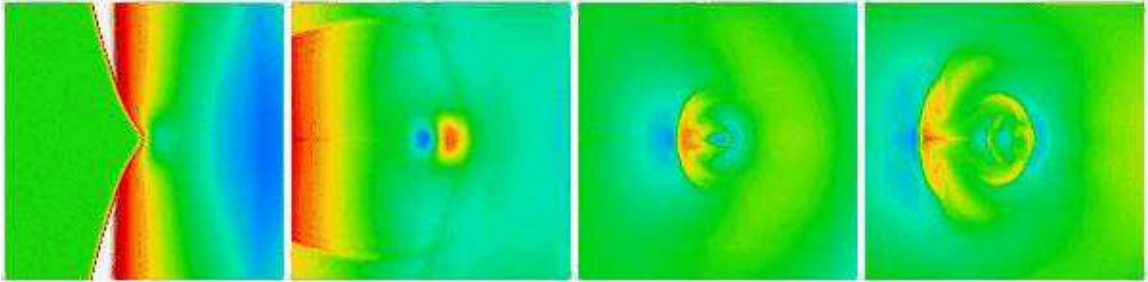}}}
\caption{Relative deviation of the gas density from the initial
value (blue color marks the regions with the density below the
initial value while yellow and red correspond to overdense
regions). The box shown is approximately 2 Mpc on a side. 
The snapshots are taken at 0.3, 0.7, 2.3, 3.9 Gyr from the
beginning of the simulation. 
\label{fig:simrm}}
\end{figure*}

\begin{figure}[ht]
\plotone{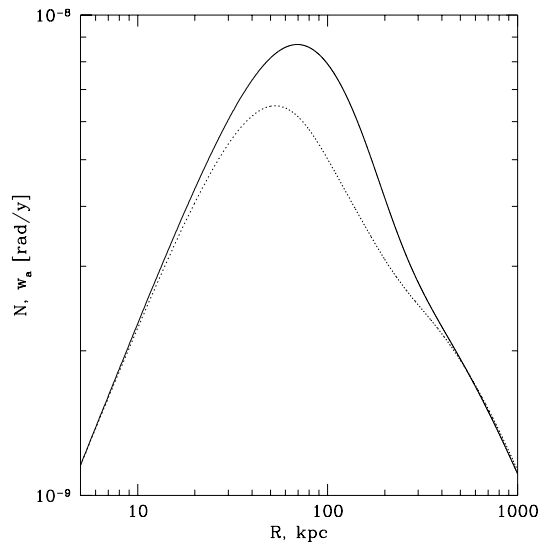}
\caption{Buoyancy $N$ (solid line)  and acoustic cutoff $w_a$ (dotted
line) frequencies as a function of radius. 
\label{fig:wbv}}
\end{figure}

Snapshots of the gas density distribution at several successive
moments of time (0, 0.3, 0.7, 1.6 Gyr from the beginning of the
simulations) are shown in Fig.~\ref{fig:simden}. Initially the plane shock
bends, when passing through 
the core, because of the higher pressure and lower temperature in the
central region. Shortly after the passage of the wave through the core, the
gas distribution is clearly elongated in the direction of motion, with
the very core regions experiencing smaller displacements than the outer
layers. The gas then begins ``sloshing'' (this term was introduced in
Markevitch, Vikhlinin \& Mazzotta 2001, see also Markevitch, Vikhlinin \&
Forman 2002) in the potential well, which in our case 
is assumed to be static. To show the structure of the density
distribution more clearly, we plot the relative deviation of the gas
density from the undisturbed value in Fig.~\ref{fig:simrm}. The first frame
corresponds to the time when the shock front is passing
through the core. Although originally the gas is swept in the
direction of the wave propagation, in $\sim$0.5 Gyr it rebounds
and produces excess emission on the opposite side of the
core. The relevant time scales for the gas sloshing are probably
set by the acoustic cutoff frequency $w_a$ and the buoyancy
(Brunt-V\"ais\"al\"a) frequency $N$, which can be written as  
\begin{eqnarray} 
w_a=\frac{c_m}{2} \frac{{\rm d ~ln} P }{{\rm ~d}~r~~~}
\label{eqn:wa}
\end{eqnarray}
\begin{eqnarray} 
N=\Omega_K\sqrt{\frac{1}{\gamma}\frac{{\rm d ~ln} s }{{\rm ~d~ln}~r~~~}},
\label{eqn:bv}
\end{eqnarray}
where $s$ is the gas entropy, $r$ is the radius, and $\Omega_K$ is
the Keplerian frequency at a given radius.  The
dependence of the acoustic cutoff and the buoyancy 
frequencies on radius (using the density and temperature profiles
according to equations \ref{ne} and \ref{te}) are shown in
Fig.~\ref{fig:wbv}. 

While strictly speaking these frequencies are
determined by the local properties of the gas (and we are interested
in global oscillations), it is likely that they set an approximately
correct timescale for the gas sloshing. Since these frequencies vary
with radius, the gas will slosh with different
periods, with the possibility of forming several asymmetric structures
at the same time on either side of the core. This is illustrated in
Fig.~\ref{fig:simrm} where the deviation of the gas density from the
undisturbed value is shown at several times. The structures
seen in the density distribution resemble the cold edges or fronts
observed by Chandra in many clusters (e.g. Markevitch et
al. 2002). The edges in this simulation are due to sloshing of the
cluster core gas and are of course transient events, which however can
persist for a few Gyr, although their shapes evolve continuously. As
usual for oscillatory motion, the gas spends most of the time near one
of the points of maximum displacement -- i.e. in the configuration
favorable for the detection of the disturbed gas as an ``edge''.

From Fig.~\ref{fig:wbv}, the shortest period of sloshing is on the
order of a Gyr. This means that it takes about a 0.5 Gyr for the gas
displaced by the initial shock to rebound to the opposite side of
the center of the potential and form an edge there. If this mechanism
is responsible for the formation of the edge in the Perseus cluster
then we can roughly estimate that the initial shock has passed through the
core some 0.5 Gyr before. On such time scales the primary wave has
enough time to propagate outside the region mapped with XMM-Newton and
the disturbed dense gas in the center has enough time to rebound.

In this interpretation the increase of gas entropy on the western side
of the cluster (see Fig.\ref{fig:press}) can be attributed to the
initial shock or sequence of shocks. If the gas in that region was
originally on the same adiabat as now observed at the eastern side of
the cluster then in a single shock interpretation the increase of
entropy by $\sim$50\% (estimated from Fig.\ref{fig:press}) would
require a shock with a Mach number of $\sim$2.6. Assuming that the
gravitating mass profile of the Perseus cluster follows the
Navarro-Frenk-White profile (Navarro, Frenk \& White, 1997) with the
parameters obtained by Ettori, De Grandi and Molendi (2002) we can
estimate that the free fall velocity reaches a Mach number of 2.6 (for
6.5 keV gas) at a distance of $\sim$300 kpc from the cluster center.
This is of course an upper limit on the velocity of the contact
discontinuity, attainable only if the infalling gas is sufficiently
dense. Depending on the geometry of the contact discontinuity the
forward shock will have a velocity similar to the velocity of the
discontinuity (e.g. bow shock) or slightly higher (for a plane parallel
shock).  Therefore the observed asymmetry in the entropy distribution
is only marginally consistent with the assumption that a single shock
was responsible for the entropy increase.  Multiple small accretion
events from roughly the same direction, which can be expected if
there is indeed a large scale filament in the direction set by the
chain of bright galaxies could more easily provide the observed rise
of the entropy.

The motion of the cD galaxy relative to the larger scale potential of
the cluster has long been suspected to play a role in clusters with
cool and dense cores (e.g. David et al. 1994, Fabian et al. 2001,
Markevitch et al. 2001). Since the distinction between the cD potential
and cluster potential may be somewhat artificial, one can think instead
in terms of perturbations of the potential in the cluster core. The
gas can then start sloshing in the disturbed potential
(e.g. Markevitch et al. 2001, 2002). We argue above that pure
hydrodynamics may cause similar sloshing. What matters is the
displacement of the gas relative to hydrostatic equilibrium, which can
be achieved either through disturbances in the gas or the gravitating
mass. In real mergers, both mechanisms are probably operating.

Markevitch et al. (2001, 2002) suggested that dissipation of the
sloshing kinetic energy may be sufficient to offset radiative gas
cooling.  Assuming that the asymmetry in the surface brightness
distribution in the inner 200 kpc region around NGC~1275 directly
characterizes the amount of sloshing energy, one can estimate this
energy as 10-20\% of the thermal energy of the gas. At a radius of
100~kpc, the cooling time is comparable to the sloshing turn around
time ($P=2\pi/w_a$ or $P=2\pi/N$) and at 200 kpc it is several times
longer than the sloshing time. Assuming that this energy is dissipated
over several sloshing turn around times (and neglecting the fact that
part of the energy would be emitted as acoustic waves which escape the
region), we conclude that the dissipated energy is not sufficient to
offset cooling. However, the magnitudes (for cooling and dissipation)
are comparable and one can imagine scenarios where the dissipation of
the sloshing kinetic energy plays a more important role. In particular
our estimate that the sloshing energy constitutes 10-20\% of the
thermal energy may be an underestimate.

\subsection{The fate of the stripped gas}
\begin{figure}[ht]
\plotone{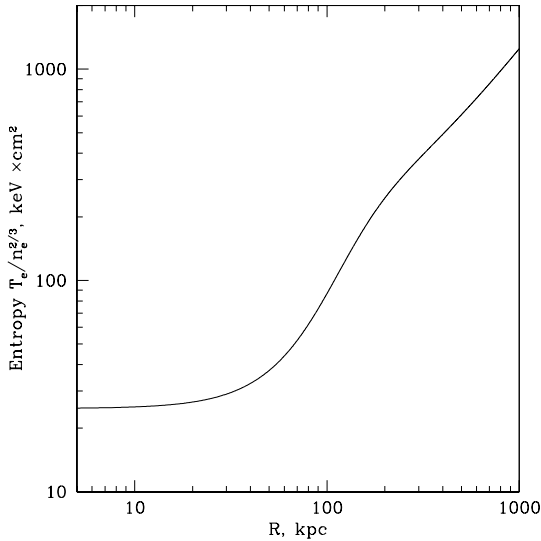}
\caption{Radial gas entropy profile calculated as $T_e/n_e^{2/3}$ using
the deprojected density and temperature profiles. 
\label{fig:entropy}}
\end{figure}

During the merger process the subcluster gas is stripped from the dark
matter potential of the infalling subcluster and mixes with the main
cluster gas. As was pointed out by Fabian \& Daines (1991), the
infalling gas does not necessarily pass through a very strong shock
during the merger process and therefore the gas may remain on
approximately the same adiabat as before the merger. This means that
even stripped and decelerated lumps of subcluster gas may continue to
move towards the center of the main cluster if their entropy is low
enough. Even neglecting mixing with the main cluster gas, this motion
can continue only until the lump reaches the radius in the main
cluster where the main cluster gas has the same entropy as the
lump. In Fig.~\ref{fig:entropy}, we plot the entropy (defined as
$T_e/n_e^{2/3}$) profile of the gas in the Perseus cluster. One can
compare the values of the entropy with the characteristic values of
the entropy found in clusters and groups (e.g. Ponman, Cannon \&
Navarro 1999; Lloyd-Davies, Ponman, \& Cannon 2000) which exceed
$\sim$100~keV~cm$^2$ even for poor groups. Thus, even if the infalling
gas does not pass through a strong shock, which would further increase
its entropy, it will find gas with comparable entropy at a distance at
least several hundred kpc from the Perseus core. The infalling gas
cannot easily penetrate below this radius. Therefore the core of the
Perseus cluster is well ``protected'' against penetration by infalling
gas (unless the infalling subcluster itself contains a cool core with
low entropy gas).

\subsection{Mergers in the absence of a dense (cool) core}

The picture discussed above assumes that the main cluster contains a
very dense (cool) core, while the infalling subcluster does not. In
reality, various combinations of these parameters are 
possible, leading to a wealth of different appearances for
mergers. In particular, if the main cluster lacks a dense (cool) core
and the entropy of the infalling gas is significantly lower than that
of the main cluster, the infalling gas may penetrate deep into the
cluster core.  The infalling cooler gas will be progressively stripped
(Fabian \& Daines 1991) from the dark matter halo and partly mixed
with the main cluster gas. Even when stripped, the gas remains
overdense compared to the main cluster gas and it will keep moving
(although more slowly) towards the bottom of the potential well. This
would result in a ``cool wedge'' penetrating deep inside (or even
crossing) the main cluster. This kind of structure is probably
observed in A1367 (Forman et al. 2003) and in Coma. Both clusters have
very high values for the gas entropy in their cores, so it is
relatively easy for infalling gas to penetrate the core. Thus the
cores in such clusters may appear much more disturbed than the cores
of the ``cooling flow'' clusters, even if the infalling subclusters
have similar parameters.

Thus if the main cluster does not have a dense and cool core, the gas
from smaller infalling subclusters may penetrate deep into the cluster
and mix with the main cluster gas. One can therefore expect the
abundance of heavy elements to be more or less uniform throughout the
cluster. However when the main cluster has a dense cool core, the
infalling gas is stopped before reaching the core and only small lumps
from any cool cores in the infalling groups themselves are ``allowed''
to penetrate deeper. These {\it infalling} cool lumps are generally
centered on a bright galaxy and are presumably also enriched in metals
compared to the bulk of the subcluster gas which is stopped at larger
distances from the center of the main cluster. Therefore this process
may serve to maintain (or even enhance) the radial abundance gradients
observed in many clusters with cool cores.
 
\section{Conclusions}

We present the first results of the 50 ksec XMM-Newton observations of the
Perseus cluster. The surface brightness and the gas temperature maps
show a wealth of substructure on all spatial scales.

The inner region ($1'-2'$) is clearly disturbed by the interaction of
the outflow of relativistic plasma from NGC~1275 with the thermal gas.
On larger scales the horseshoe structure of the gas temperature and
the surface brightness structure suggest an ongoing merger along the
direction defined by a chain of bright (optical) galaxies. The gas
from the infalling subcluster may be stripped from the dark matter
halo at a large distance from the core of the Perseus cluster, with
only compression/rarefaction waves passing through the core of the
main cluster. These waves may induce ``sloshing'' in the gas in the
cluster core. As a result an ``edge'' (or even several edges) in the
surface brightness and in the gas temperature distribution may appear.

Similar structures may be present in other clusters with peaked (cool)
cores which are presently accreting smaller subclusters
(e.g. Centaurus cluster). For clusters without a cool dense core, the
situation may be different since the subcluster gas is cool and can
penetrate deep inside the core of the main cluster and form a cool
wedge of partially stripped and mixed gas (e.g. Coma, A1367).

\acknowledgments

We thank Jean Ballet, Michael Freyberg, Marat Gilfanov, Sebastian
Heinz, Nail Inogamov, Francesco Miniati, Ewald M\"uller and Rashid
Sunyaev for valuable discussions.  W. Forman and C. Jones thank MPA
for its hospitality during their summer 2002 visits and acknowledge
support from the Smithsonian Institution and NASA (NAS 8 39073 and NAG
5 9944).  We acknowledge use of the Digital Sky Survey prepared by ST
ScI (NASA grant NAG W-2166) from photographic data obtained with
support from the National Geographic Society to CIT.

This work is based on observations obtained with XMM-Newton, an ESA
science mission with instruments and contributions directly funded by
ESA Member States and the USA (NASA).

\label{lastpage}

\end{document}